\documentclass[aps,11pt,superscriptaddress,nofootinbib,notitlepage,showkeys]{revtex4-2}

\pdfoutput=1
\usepackage[T1]{fontenc}
\usepackage{comment}
\usepackage{graphicx}
\usepackage{bm,epsfig,ifthen}
\usepackage{amsfonts,mathtools,amssymb,amsmath,stmaryrd}
\usepackage{color}
\usepackage[breaklinks]{hyperref}
\usepackage{epstopdf, epsfig}
\usepackage{natbib}
\usepackage{grffile}
\usepackage{amsthm}
\usepackage{bbm}
\usepackage{MnSymbol}
\usepackage{algorithm,algorithmic}

\graphicspath{{Figs/}}

\newcommand{\de}{\mathrm{d}}

\newcommand{\mU}{{\mathcal U}}

\newcommand{\bk}{{\bf k}}

\newcommand{\be}{\begin{equation}}
\newcommand{\ee}{\end{equation}}
\newcommand{\bsp}{\begin{split}}
\newcommand{\esp}{\end{split}}

\newcommand{\bi}{\begin{itemize}}
\newcommand{\ei}{\end{itemize}}

\newcommand{\rroc}{\color{black}}

\newcommand{\hs}{\hspace{0.2cm}}

\newcommand{\cm}{\color{black}}
\newcommand{\cb}{\color{black}}
\newcommand{\xdam}{{N-DAM}}

\graphicspath{{figs/}{newfigs/}}

\begin{document}

\title{\cm Blowup  driven by critical balance in  a differential kinetic model of gravity wave turbulence}

\author{Daniel Schubring}
\affiliation{Initiative for the Theoretical Sciences, The Graduate Center, CUNY, 365 Fifth Ave, New York, NY 10016, USA}
\author{Vladimir Rosenhaus}
\affiliation{Initiative for the Theoretical Sciences, The Graduate Center, CUNY, 365 Fifth Ave, New York, NY 10016, USA}

\author{Simon Thalabard}
\affiliation{Institut de Physique de Nice, Universit\'e C\^ote d'Azur CNRS - UMR 7010, 17 rue Julien Laupr\^etre, 06200 Nice, France}

\date{\today}
\begin{abstract} 
\cm
We describe the blowup scenarios in a phase-parametrized differential approximation kinetic model (\xdam), inspired by the physics of deep water surface gravity waves and  recently obtained using  large-$N$ summation techniques  under a local approximation in wavenumber space.
Previous work showed that  the model admits   steady-state solutions interpolating between the Kolmogorov-Zakharov spectrum $E(\omega)\propto \omega^{-4}$ and   either a strong-turbulence regime $E(\omega)\propto \omega^{-2}$ or the  Phillips critical-balance spectrum $E(\omega) \propto \omega^{-5}$     at small scales. 
These solutions reproduce scaling regimes expected in gravity-wave kinetics, 
suggesting that the  \xdam\, may serve as an effective augmented version of  an earlier  differential approximation model  introduced  by Hasselmann. 
Here we investigate  finite-time blowup in the \xdam\, and show that it is generically governed by the critical-balance regime. This leads to a non-Kolmogorov finite-time transfer of the  energy from the IR towards the UV for any value of the parameter $\phi \in [0,\pi)$. 
We observe a bifurcation in the blowup dynamics from continuous to discrete  self-similarity as  $\phi$ is increased above a critical value $\phi_*\simeq 2.7$. To our knowledge, this is the first example of a discretely self-similar blowup in the kinetic theory of waves.  
\cb
\end{abstract}

\maketitle
\tableofcontents
\section{Introduction}
The nonlinear interactions  of  gravity waves (GW) on the deep ocean  surface  provide a classical testbed for the theory of wave turbulence \cite{hasselmann1962non,badulin2005self,korotkevich2007numerical,korotkevich2019dissipation,zakharov2012generalized}. The theory is  formalized by the (homogeneous) Hasselmann wave kinetic equation \cite{hasselmann1962non,zakharov2010energy} for the wave density $n({\bf k},t)$, 
\be
	\partial_t n({\bf k},t)  =S_{nl} + S_{f}+ S_{diss}
\ee
describing   cascading processes   mediated by resonant quartets of modes in the collision integral $S_{nl}$,
in the presence of forcing $S_{f}$ and damping $S_{diss}$.
 The nonlinear term $S_{nl}$ prescribes  the standard phenomenology of wave turbulence, supporting both thermal Rayleigh-Jeans (RJ) and non-thermal Kolmogorov-Zakharov (KZ) scaling regimes --- see, e.g. \cite{newell2011wave, nazarenko2015wave,zakharov2019weak} and references therein. In particular, the KZ solution associated to the 
 direct cascade of kinetic energy 
\be
 	E(\omega,t) = \int_{\mathbb R^2} d {\bk}\,  \delta\left( \omega(\bk) - \omega\right) \omega(\bk) n({\bf k} ,t),\quad  \omega(\bk) =\sqrt{ g |\bk|}
\ee  
produces the scaling  $E(\omega) \propto \omega^{-4}$,  consistent with oceanic observations---
the energy representing the fluctuations of the surface elevation.
 While one could expect the KZ spectrum to hold from kilometric  down to  capillary scales, oceanic field measurements \cite{badulin2007experimental, lenain2017measurements,zakharov2017balanced} rather support the idea that  KZ  breaks down at metric scales \cite{lenain2017measurements},  transitioning into a regime  $E(\omega) \propto \omega^{-5}$ known  as the Phillips spectrum \cite{phillips1957generation, newell2011wave, nazarenko2015wave,zakharov2012generalized, lenain2017measurements}.

\cm One interpretation ties  the Phillips spectrum to   wave-breaking mechanisms and singularity formation, in particular, white-capping events appearing under strong wind forcing \cite{dyachenko2016whitecapping,korotkevich2019dissipation,kuznetsov2004turbulence}. \cb This  leads to a balance  between non-linear energy transfers and suitably parametrized dissipation mechanism \cite{zakharov2012generalized,badulin2020phillips}.  This is a form of \emph{critical balance}, which, more generally, denotes a conjectured regime in which the state adjusts itself so that the growth of the nonlinear term  saturates and becomes of the same order as the linear term \cite{Goldreich,Newell,NS}.  Another viewpoint interprets the Phillips spectrum as a form of  statistical restoring of scaling symmetry  \cite{NS,newell2011wave}---akin to the  statistical restoring of symmetries in the classical theory of  Navier-Stokes turbulence \cite{frisch1995turbulence, MailybaevTransactions2022}. 
\cb
\cm
The scope of our work is  to  discuss an effective kinetic model of gravity wave kinetics in which the critical-balance regime can be studied mathematically, and  drives a singularity  in the kinetic equation.
\cb

While one can compute the kinetic equation perturbatively in the nonlinearity \cite{RS1, RS2, RSSS, schubring2023fokker, HuRose, FR1}, reaching the strong turbulence regime requires more. \cm
Insights can be obtained by considering the full kinetic equation---valid at all scales---by taking  a large-$N$ limit, formally consisting of vectorializing the underlying field into a large number of components. 
This approach, which can be thought as a form of mean-field limit,  is perhaps most familiar in the turbulence literature in the context of the random coupling model \cite{Kraichnan, Betchov, Hansen, eyink1995large}---in the large-$N$ limit the model is solvable and the direct interaction approximation is exact. 
\cb
A further simplification occurs if in addition to large-$N$, one instead takes  an interaction that is strongly local in momentum space. Such a  large-$N$  differential approximation model  (DAM)  was introduced in \cite{rosenhaus2024strong}, where it was found that  the collision integral is a fourth order differential operator,  yielding the  non-linear conservation law  in frequency space 
\be
	\label{eq:RSDAM}
	\begin{split}
	& \partial_t E (\omega,t) +\partial_\omega P =  F+D ,\quad P =- \omega^2\partial_\omega\left(\omega^{-1} K \right)~,\\
  	& K(\omega,t)= \frac{\omega^{10} E^{4} \partial^2_{\omega \omega} (\omega^4 E^{-1})}{|1-e^{j\phi}U |^2},\quad U=\omega^{10}\partial_\omega \left(E\omega^{-4}\right).
	\end{split}
\ee
Here the powers of $\omega$ have been fixed to those appropriate for the gravity wave case, and the units have been rescaled such that the only free parameter in the model is a phase $\phi$ ranging from $0$ to $\pi$---see Appendix \ref{Appendix A} for further details. We emphasize that (\ref{eq:RSDAM}) is not contrived to give any particular scaling in the non-weak (strong) turbulence regime
\cm but follows from summing a perturbation series to all orders in the interaction. 
\cm
In particular, the phase parameter $\phi$ is  not introduced to  reproduce a predetermined spectrum.
Rather, it parametrizes a family of closures that interpolate between distinct dynamical regimes of the effective kinetic equation while preserving its general structure.
\cb

We later refer to 	the model (\ref{eq:RSDAM}) as the \xdam.\, 
At this stage,  we note  that formally setting the function $U$  to zero reduces the \xdam\,  to the isotropic version of  a diffusion approximation already found by Hasselmann et al \cite{hasselmann1985computations}, written  in a general form by Dyachenko et al.  \cite{DyachenkoNewellPushkarevZakharov1992}, and  studied  in greater detail  by Pushkarev \& Zakharov \cite{zakharov1999diffusion} in the specific context of gravity waves (see also \cite{nazarenko2006sandpile, nazarenko2015wave}). The model then 
 admits pure scaling steady states in terms of KZ and RJ solutions.
The presence of nonzero $U$ alters the nature of the steady states. In particular, the  \xdam\,  supports  steady-state direct cascade solutions which interpolate between  Kolmogorov-Zakharov spectrum $E(\omega) \propto \omega^{-4}$ in the infrared (IR) and a  strong-coupling regime $\propto \omega^{-2}$ in the ultraviolet (UV). 
In addition, at the specific value of $\phi=\pi$, the model also supports a transition from KZ to  the Phillips regime $\propto \omega^{-5}$.
\cm One could object that  neither of the two simplifying features used to derive the \xdam\, should \emph{a priori} be  valid for gravity waves: there is no physically natural  multicomponent extension of  the ocean surface, and the interaction of a quartet of gravity waves is not expected to be  local in momentum space. 
Our intent, however,   is not to claim physical realizability, but to use the combination of a large-$N$ limit and a diffential approximation as a controlled setting in which a kinetic model of gravity waves  
can be studied rigorously, that is not restricted to the weakly nonlinear regime.
  In this perspective, we will find that the model (\ref{eq:RSDAM})  has two novel and surprising dynamical features,
which connect to the question of finite-time blowup in kinetic equations.

\cb 
Finite capacity cascades --- such as the ones for gravity waves --- are characterized by a traveling front reaching infinite frequency in finite time. The spectrum that is left behind is usually regarded as a non-universal transient, which is replaced at a later time by the stationary solution resulting from a second stage of backward propagation from infinite to finite frequency.  
Here, the first feature  we report is that the front  leaves, nontrivially,  Phillips scaling in its  trail. 
 In other words,  critical balance  turns out to be the driving mechanism for the finite-time blowup. It prescribes a finite-time propagation of the energy to the UV, converging to  an asymptotic similarity solution with  Phillips scaling. This scenario holds for any value of the parameter $\phi$, so it has no connection to the late-time stationary Phillips scaling, which only holds for $\phi = \pi$.  \\
We find a second novel feature: The similarity solution bifurcates from  continuous to discrete (or periodic) as the phase parameter $\phi$ increases towards $\pi$, at a critical phase of $\phi_\star\simeq 2.7$. Although similarity solutions have been studied in many families of DAM associated to wave kinetics, the \xdam\, provides  the first explicit example of a non-continuous similarity blowup in the context of the kinetic theory of wave turbulence---to the best of our knowledge. 

The paper is organized as follows. \S \ref{sec:0} reviews the basic phenomenology of the \xdam\, and its various steady-state solutions,
\S\ref{sec:1} describes the (time-dependent) asymptotic similarity and the interplay between the Phillips, weak and strong-coupling regimes in the blowup solutions. \S\ref{sec:3} describes  a bifurcation of the blowup from self-similar to periodic as $\phi$ is varied.
\S\ref{sec:4} formulates concluding remarks.

\section{Steady-state similarity}
\label{sec:0}
A detailed discussion on the steady states produced by the \xdam~\eqref{eq:RSDAM} can be found in \cite{rosenhaus2024strong}; here we recall the basic phenomenology. The magnitude of the function $U$ in the denominator provides an  interpolation between the Pushkarev-Zakharov (\emph{weak}) DAM---obtained in the limit $U\rightarrow 0$---and a \emph{strong} DAM in the opposite limit $|U|\to \infty$. The weak and strong asymptotics are defined   through the substitutions  $K\mapsto$
\be
 K_W= \omega^{10} E^{4} \partial^2_{\omega \omega} \left(\omega^4 E^{-1}\right) \hs \text{(weak)} \quad \ \ \ \ \  \text{or}\quad  \ \ \ \ \
  K_S:=\omega^{-10} E^{4} \dfrac{\partial^2_{\omega \omega} \left(\omega^4 E^{-1}\right)}{\left[\partial_{\omega} \left(\omega^{-4} E\right)\right]^2} \hs\hs  \text{(strong)}
\label{eq:K0-oo}
\ee
Both  models admit  a scaling solution with constant  energy flux $P_0>0$, respectively:
\be
	\label{eq:KZ-WS}
	E_W = \left(\dfrac{P_0}{56}\right)^{1/3}\omega^{-4}  \quad  \text{or}  \quad	E_S =  \dfrac{6}{5} P_0  \,\omega^{-2}~,
\ee
associated to direct cascades of energy, transferring energy form small to large frequencies (i.e. from large to small scales). The solution $E_W$ is the KZ scaling.\\
\begin{figure}
\begin{minipage}{0.49\textwidth}
\includegraphics[width=\textwidth]{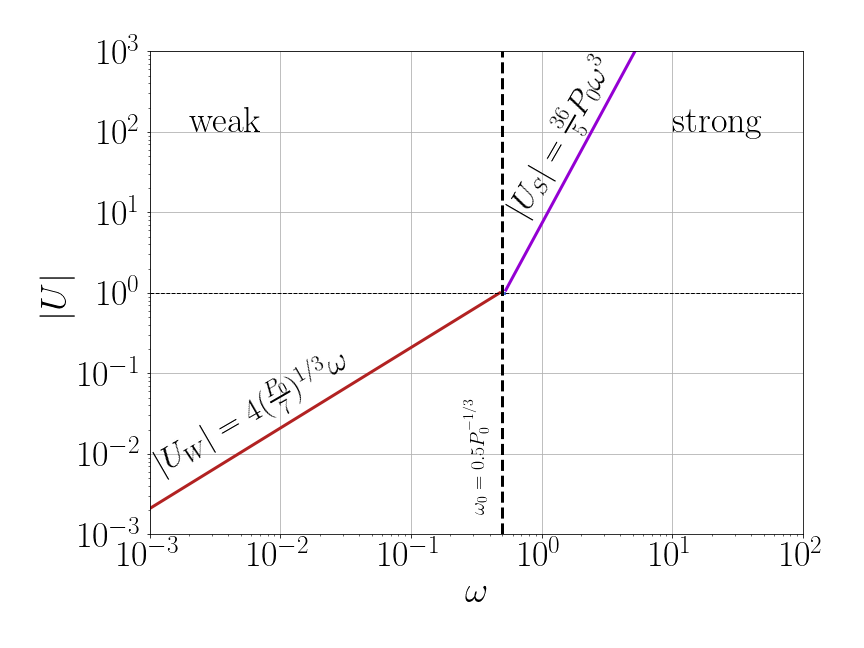}
\end{minipage}
\begin{minipage}{0.5\textwidth}
\includegraphics[width=\textwidth,trim=0cm 0cm 0cm 0cm,clip]{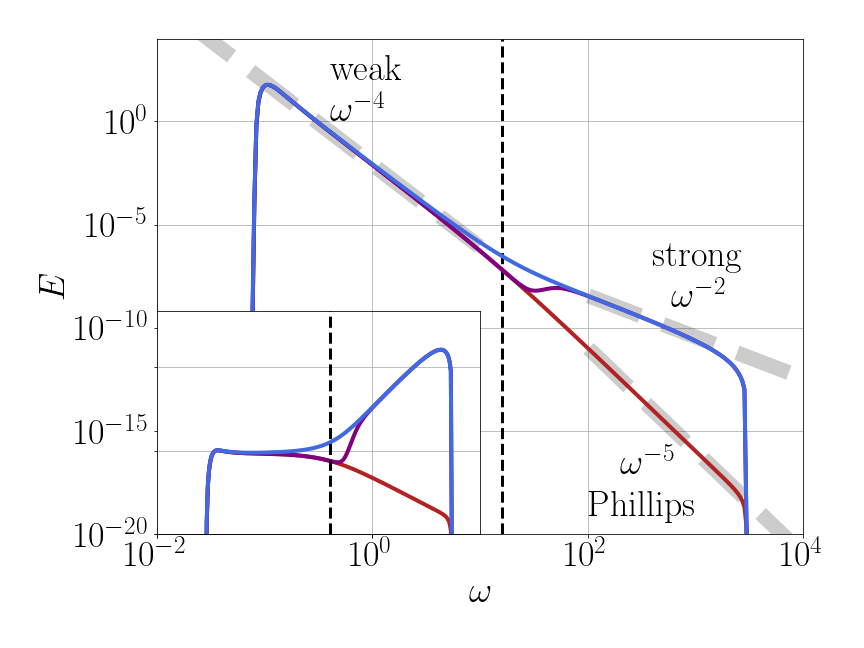}
\end{minipage}
\caption{
Left : Cartoon phenomenology of the direct cascade when $\phi \in [0,\pi)$.  Right:  Steady-state energy spectra observed in numerical simulations with forcing at $\omega \simeq 10^{-1}$, small-scale dissipation at $\omega> 2\times 10^3$, and large-scale damping at $ \omega<8\times 10^{-2}$ for $\phi = 0, 0.99\pi, \pi$ in blue, purple, red respectively. The asymptotic scaling solutions $E_{W}, E_{S}, E_{P}$ are indicated by dashed lines. \cm The inset shows the spectra compensated by the weak KZ solution $\propto \omega^4$. The dash vertical lines show the value $\omega_0 = 0.5 P_0^{-1/3}$, indicative of the weak to strong transition.
\cb}
\label{fig:1}
\end{figure}

 For finite $U$, the steady state solutions interpolate between the weak and strong scaling regimes in the IR and UV, respectively. This is shown in the right panel of Fig.~\ref{fig:1}, which plots the results of direct numerical simulations of the DAM with prescribed large-scale forcing and dissipation. For $\phi=0$, the steady-state solution for $E(\omega)$  interpolates directly between the two scaling solutions of 	Eq.~\eqref{eq:KZ-WS}. Heuristically, the transition from  the weak to the strong regime occurs when the function $|U|$ becomes of order $1$, which happens when  $\omega \simeq  \omega_0 =\frac{1}{2}P_0^{-1/3}$.  
Note that for values of $\phi$ near $\pi$, the solution develops a bump-like feature near the weak to strong transition (Shown in Fig.~\ref{fig:1} for $\phi=0.99\pi$).
\rroc

Precisely at $\phi=\pi$,  a second asymptotic scaling regime is possible in the UV \cite{rosenhaus2024strong}. This is the Phillips spectrum characterized by 
\be
E_{P}= \frac{1}{9}\,\omega^{-5}.
\ee
In principle, a solution interpolating to strong scaling $E_S$  is also  possible, but the Phillips UV behavior appears more robust.
The  numerics in Fig.~\ref{fig:1} show that the Phillips UV asymptotics indeed emerge in  forced-dissipated numerical simulations of the \xdam. 

Note that the pure solution $E_P$ does not depend on the flux $P_0$, and a direct substitution in the \xdam~ \eqref{eq:RSDAM} would cause the denominator in $K$  to diverge. This is not fundamentally problematic:
 the actual steady-state solution $E$ is never exactly $E_P$, as it  includes $P_0$-dependent sub-leading terms which vanish faster than $\omega^{-5}$ and  prevent any true divergence at finite $\omega$. However, we point out that the realization of the Phillips spectrum requires some numerical care.  In particular, a precise tuning of the viscosity is required, in order to prevent numerical instabilities caused by  the small values taken by the denominator. 

We will next show that  Phillips scaling arises for generic values of $\phi$ in the finite-time inviscid evolution of the \xdam\, toward  small scales, describing the finite-time blowup occurring in the  limit $\nu\to 0$.

\section{Time-dependent similarity}
\label{sec:1}
Steady-state  solutions have finite capacity if the integral $\int E(\omega) d\omega$ is convergent, which we expect to be the case if the solution has  the strong scaling  $\propto \omega^{-2}$ in the UV. A common heuristic  predicts that finite capacity  requires the finite-time propagation of any  perturbation   down to the UV, generically associated  to the finite-time blowup of a vorticity norm~\cite{nazarenko2015wave, thalabard2015anomalous}.
The simplest blowup scenario is mediated by self-similar intermediate asymptotics \cite{barenblatt1972self}. In other words, the solution  converges  in finite time towards a universal self-similar profile, characterized by a sharp UV front leaving in its trail an anomalous scaling. 
This  has been well-studied for the usual nonlinear diffusion models associated to  wave turbulence kinetics \cite{giga1987characterizing,diez1992selfa,diez1992selfb,connaughton2004warm,thalabard2015anomalous,thalabard2021inverse}, and
we point out that the analysis can be extended to discrete cascade models where the blowup scenarios are known to be richer~\cite{mailybaev2012renorm,mailybaev2013bifurcations,campolina2025non}.
\rroc

In this section we  extend the analysis of the \xdam\, to such time-dependent settings and analyze  the self-similar blowups  individually supported by  each of three regimes:  $\mathcal{U}=0$,  $|\mathcal{U}| \to \infty$,  and $0<|\mathcal{U}|<\infty$.  The interplay between the different  blowups will be discussed in \S \ref{sec:3}.

\subsection{Weak case:  $U=0$}
\label{ssec:31}
The most direct way to capture the self-similar blowup is to introduce the ansatz 
\begin{equation}
	 \label{eq:ansatz}
	E(\omega, t) = (t_* - t)^{bx}  \omega^{-11/2} \Omega(\eta), \quad \eta(\omega,t) = (t_* - t)^{b} \omega~.
\end{equation}
For positive exponent $b>0$, the ansatz describes a self-similar solution with a sharp  front $\omega_* \propto (t_*-t)^{-b}$ propagating towards the UV. The scaling function $\Omega(\eta)$ depends on the similarity variable $\eta= \omega/\omega_*$.  It decays to $0$ smoothly in the UV  ($\eta\gg 1$)  and  has asymptotic algebraic scaling behavior  $\Omega \propto \eta^{-x}$ ($\eta\ll 1$) in the IR. This translates   into $E(\omega) \simeq \omega^{-x-11/2}$ sufficiently far from the front.  
 Setting $U=0$ and \rroc substituting this ansatz into the \xdam\, \eqref{eq:RSDAM} sets
\begin{align}
	\label{eq:b}
	b(x)=-\frac{1}{2x}~.
\end{align}
This prescribes that the profile $\Omega$ satisfy a $4^{th}$-order ordinary differential equation (ODE),  which we compactly write as 
\be
	2xD_{-11/2}D_{-9/2} \left(\Omega^4 D_{17/2} D_{19/2} \Omega^{-1}\right)-D_x\Omega =0~,
	\label{eq:ODEweak}
\ee
in terms of the differential  operators $D_\alpha$ defined through
\be
	\label{eq:Dalpha}
	D_\alpha f = \eta \dfrac{\de}{\de \eta} f + \alpha f  \quad \text{ for } \alpha \in \mathbb R~.
\ee
The derivations of Eq.~\eqref{eq:ODEweak}-\eqref{eq:Dalpha} can be found in  Appendix~\ref{sec:technical}. 

\begin{figure}
	\includegraphics[width=0.45\textwidth]{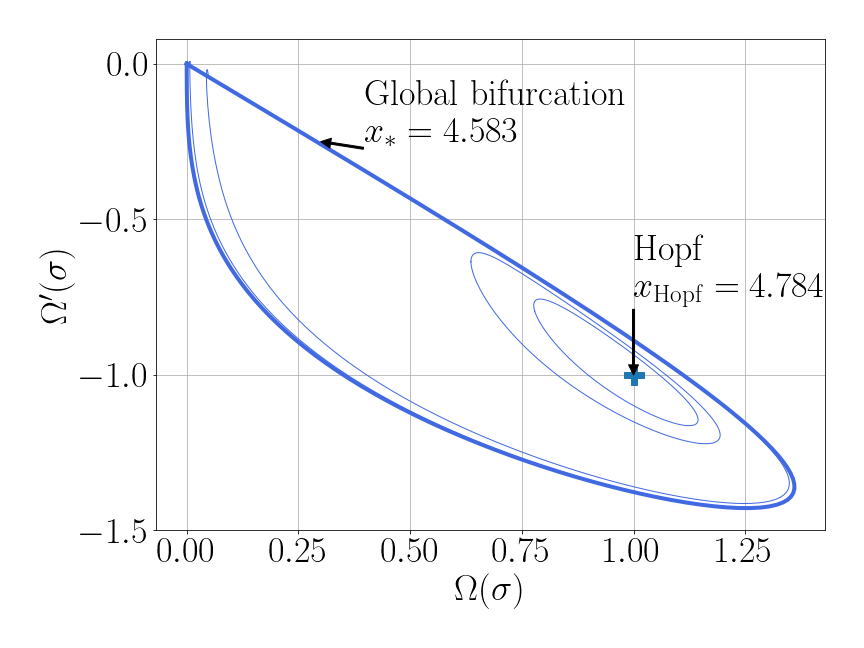} \ \ \ \ 
	\includegraphics[width=0.45\textwidth]{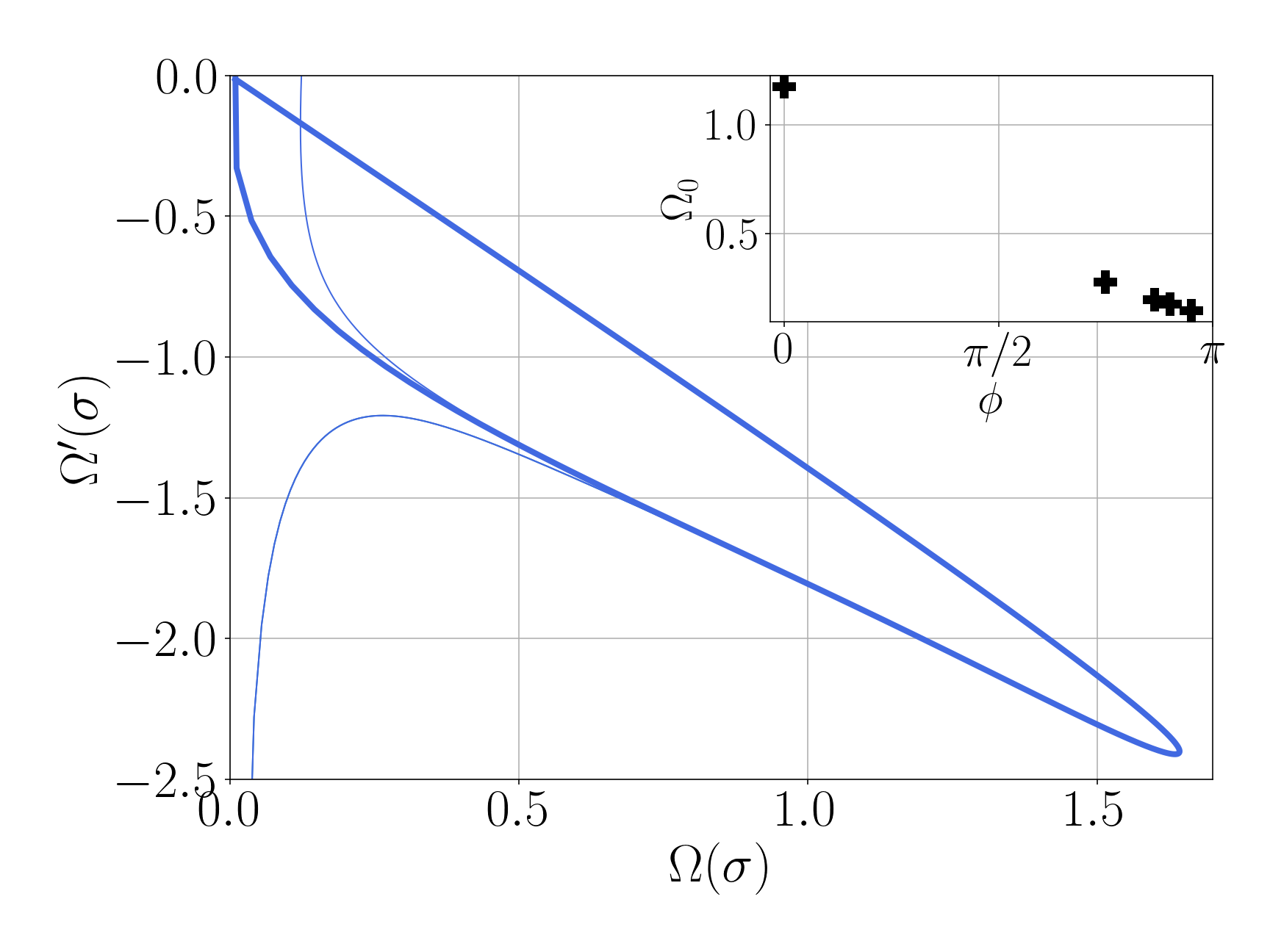}
		\includegraphics[width=0.45\textwidth]{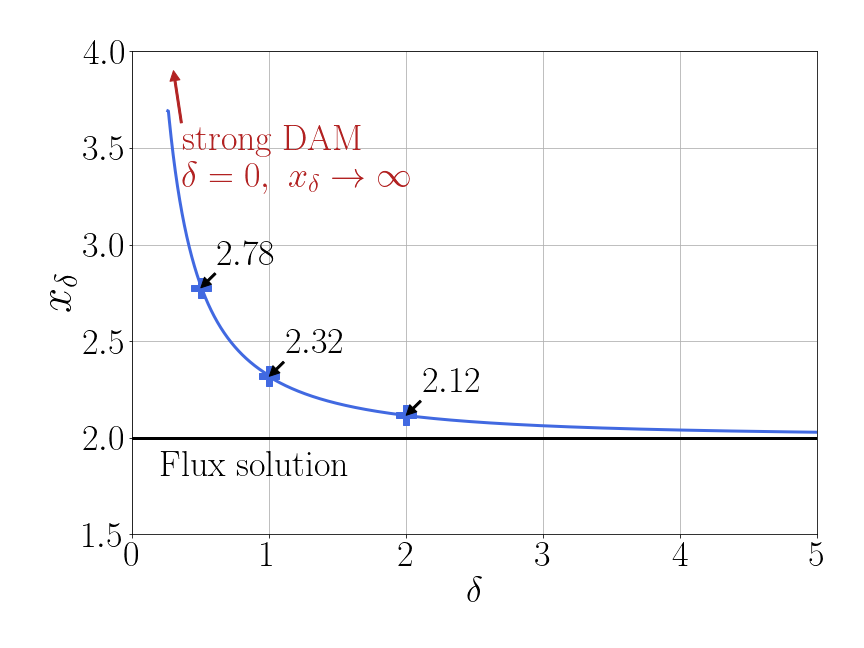}
		\caption{Top left: Numerical continuation scheme in the weak regime:
The similarity profile $\Omega$  is identified as a global bifurcation of \eqref{eq:ODEweak} and tracked from the  numerical continuation of a branch of limit cycles initiated from a Hopf bifurcation.  Here, both axes are rescaled so that the Hopf point is at $(1,-1)$. Top right:  Phase portrait of the similarity profile at $\phi=3\pi/4$ in the generic case, obtained from a shooting method --- see text. Inset shows the Phillips coefficient $\Omega_0$ for 5 different values of the phase parameter. 	Bottom: The anomalous exponent $x_\delta$ as a function of the nonlinearity exponent $\delta$ in the $\delta$-perturbed strong regime. 
}
	\label{fig:3}
	\end{figure}

For any choice of $x$, the ODE \eqref{eq:ODEweak} has a solution obeying $\Omega \sim \eta^{-x}$ in the IR, but only for  particular choices of $x$ (possibly non -unique) will $\Omega$ also have the appropriate boundary conditions in the UV involving  $\Omega$ decaying faster than any power law as $\eta \to \infty$. In that sense, Eq.~\eqref{eq:ODEweak} represents  an eigenvalue problem.
Changing variables $ \eta \to \sigma = \log \eta$,  the differential operators become $D_\alpha \to  \frac{\de}{\de \sigma} +\alpha \, \text{Id} $, and this transforms \eqref{eq:ODEweak} into an autonomous system:
Appropriate choices for $x$ may then be found numerically using a variety of techniques, from  brute-force optimization \cite{campolina2025non} to shooting methods \cite{dombre1998intermittency} and continuation techniques \cite{thalabard2021inverse}. Here we used the latter  to find  a unique value $x \simeq -0.917$ leading to the scaling law $E  \simeq \omega^{-4.583}$ for $ \omega \ll \omega_*$ (see the left panel of Fig.~\ref{fig:3}).  We later refer to the value $4.583$ as the weak anomalous scaling exponent.

Let us  comment on our notation.   The similarity solution prescribes in particular a blowup at the level of  the  norms defined as 
\be
	\label{eq:bkm}
	|W|_p= \left(\int_0^\infty W^{p} d\omega\right)^{1/p},\quad W(\omega,t) =  \omega^{11/2}E(\omega,t)~.
\ee
Using Eq.~\eqref{eq:ansatz}-\eqref{eq:b},  a direct calculation yields the diverging behavior
$|W|_p \propto (t_*-t)^{-\frac{1}{2} \left(1+\frac{1}{xp}\right)}$. The  blowup rate becomes independent of the IR similarity exponent $x$ in the large-$p$ limit, $|W|_\infty \propto (t_*-t)^{-\frac{1}{2}}$.  This is reminiscent of the Beale-Kato-Majda criterion tying blowup in the incompressible Euler equations  to the vorticity supremum diverging  at least as rapidly as $(t_*-t)^{-1}$\cite{beale1984remarks,bustamante2012interplay}. This analogy motivates our choice of notation $W$ and $\Omega$, suggestive of a vorticity.
\subsection{Strong case:  $|U| \to \infty$}
\label{ssec:strong}
We recall that the strong regime asymptotics  corresponds to setting   $K \to K_S$,  as prescribed by Eq~\eqref{eq:K0-oo} in the \xdam\, \eqref{eq:RSDAM}. The strong DAM supports the finite-capacity KZ scaling $E \propto \omega^{-2}$. However, it does not support a finite-time blowup.
At a heuristic level, this apparent contradiction comes from the energy flux $P = -\omega^2 \partial_\omega (\omega^{-1}K_S)$  being  homogeneous of order 1 in $E$, meaning that changing $E \to \lambda E$ entails $P \to \lambda P$ for any $\lambda >0$. While it is not clear if this can be  explicitly mapped to  standard diffusion,  we  expect to see a diffusive dynamics---akin to the heat equation, and with no sharp-front solutions.

This heuristics can be substantiated from a perturbative argument. To this end, let us slightly perturb the strong regime into
\be
	\label{eq:delta}
	 K_S \to K_S^{(\delta)}:=  \omega^{2\delta-10} E^{4+\delta} \dfrac{\partial^2_{\omega \omega} \left(\omega^4 E^{-1}\right)}{\left[\partial_{\omega} \left(\omega^{-4} E\right)\right]^2},\quad \delta >0~.
\ee
Namely, we impose a small nonlinearity in the energy flux $P[\lambda E,\omega] = \lambda^{1+\delta} P [E,\omega]$, such that the resulting $\delta$-perturbed \xdam\, shares the RJ solutions  and energy flux pure scaling solutions $\propto \omega^{-2}$ of the strong unperturbed case ($\delta=0$). 

The $\delta$-deformation trick allows us to find a unique self-similar solution,  just as in the weak limit. Following the steps of \S\ref{ssec:31}, the relevant vorticity turns out to be  $W^{(\delta)} = \omega^{2+\frac{1}{\delta}} E$, which  blows up at  the rate
$|W^{(\delta)}|_\infty \propto (t_*-t)^{-1/\delta}$. 
In the strong limit  $\delta \to 0$, one needs to compensate the energy spectrum by a diverging scaling exponent ${2+\frac{1}{\delta}} \to \infty$. This implies that in the strong limit the energy profile is smooth, decaying faster than any power law. This is  reflective of  diffusive behavior and the absence of a self-similar solution blowing up in finite time.

As a consistent illustration, the bottom panel in Fig.~\ref{fig:3} shows  an estimate of the anomalous similarity exponent  as a function of the nonlinearity parameter $\delta$, obtained from numerical continuation. The increasing deviation from the constant-flux exponent $2$ as $\delta \to 0$ is consistent with a diverging behavior $x_\delta \to \infty$, indicative of the diffusive behavior of the strong regime.

\subsection{Generic case: $0<|U|<\infty$}
\label{ssec:generic}
The analysis performed in the weak case extends  to the generic case of finite and nonzero $U$.  
Inserting the ansatz \eqref{eq:ansatz}-\eqref{eq:b} into Eq.~\eqref{eq:RSDAM}  additionally prescribes $x=-1/2$, 
in order to get a  time-independent equation for the similarity profile $\Omega(\eta)$. A few lines of algebra then also prescribe that it be solution of the fourth-order dynamics
\be
	0=D_{-1/2} \Omega +D_{-11/2}D_{-9/2}\left( \dfrac{\Omega^4 D_{17/2}D_{19/2} \Omega^{-1}}{\left|1-e^{j\phi}\mathcal V\right|^2}\right),\quad 
\mathcal V= \eta^{-1/2}D_{-19/2} \Omega~.
\label{eq:ODEPhil}
\ee
Such   similarity solutions prescribe a blowup $|W|_\infty \propto (t_*-t)^{-1/2}$, as in the weak case.  The parameter $b$ \eqref{eq:b} takes the value of $1$, so the fronts propagate as $\omega_* \propto (t_*-t)^{-1}$ and  leave behind the Phillips spectrum: $W \propto \omega^{1/2}$ or  equivalently $E \propto \omega^{-5}$ for $\omega \ll \omega_*$. For clarity, we emphasize that while the Phillips scaling is occurring  well behind the edge of the front, at $\omega\ll \omega_*$, it occurs in a non-weak regime of large enough $\omega$ so that $|U| \gtrsim 1$, as we will see in the next section.

A key difference between the generic equation \eqref{eq:ODEPhil} and the weak equation \eqref{eq:ODEweak} is that the weak equation may be put in an autonomous form simply by transforming to a variable $\sigma=\log \eta$, and thus it is symmetric under translations in $\sigma$, or equivalently rescalings of $\eta$. This means that associated to any solution asymptotically scaling as $\Omega(\eta)\sim \Omega_0\eta^{-x}$ there is actually a whole family of solutions related by rescalings of $\eta$, and thus involving different values of the coefficient $\Omega_0$. The natural eigenvalue problem in the weak case is finding the value of $x$ such that $\Omega(\eta)$ has the appropriate boundary conditions, and $\Omega_0$ is not relevant information.

However, in the generic equation \eqref{eq:ODEPhil}  the translation symmetry is explicitly broken by the $\mathcal{V}$ term. Now there is no longer a family of solutions related by rescalings, and the coefficient $\Omega_0$ is crucial information. Even though the value $x=-1/2$ is fixed, there is still an eigenvalue problem involving selecting the value of $\Omega_0$ such that $\Omega(\eta)$ has the appropriate boundary conditions.

The eigenvalue problem may be solved numerically for any value of $\phi<\pi$, see the top right panel of Fig.~\ref{fig:3}. The numerical method used is a variant of a shooting method. Fixing a trial value for $\Omega_0$, the asymptotic solution to \eqref{eq:ODEPhil} may be used to set approximate initial conditions in the IR. These initial conditions are tuned to increasingly greater precision to balance between two classes of UV asymptotic behavior, and the light blue curves in Fig.~\ref{fig:3} reflect solutions with slightly imprecise initial conditions. The solution constructed in this manner will eventually reach a point $\sigma$ at which either $\Omega(\sigma)=0$ or $\Omega'(\sigma)=0$, depending on whether the trial value for $\Omega_0$ was set too high or low, respectively. At the exact value of $\Omega_0$ which solves the eigenvalue problem, both conditions $\Omega(\sigma)=\Omega'(\sigma)=0$ should hold at the same point $\sigma$, and this value may be bounded from above and below to increasingly greater precision.

\section{Blowup scenarios}
\label{sec:3}

We now analyze the interplay of blowups between the three regimes. Our  main finding  is  that 
the blowup in the \xdam\, is  determined not by  the strong DAM,  but by the  Phillips regime. In fact,  the blowup asymptotic solution precisely produces Phillips scaling $\propto \omega^{-5}$---meaning that critical balance extends across  the entire UV range. Through numerical simulations, we show that   convergence towards the self-similar profile can proceed via either a  $weak\to Phillips$ or $strong\to Phillips$ transition, depending on the magnitude of the initial conditions. This behavior is robust for most values of  $\phi$; however,  as $\phi$ approaches $\pi$, we give evidence for a transition in the blowup structure, from  self-similar to periodic .

\subsection{Convergence to Phillips similarity}
When starting from an initially smooth (say Gaussian)  profile sharply localized around $\omega_i=O(1)$  
and with energy $E=O(1)$,   the interplay between the various similarity regimes is  driven by the magnitude of the  $\mU$ term, as discussed for steady-states in \S \ref{sec:0}.

Phillips similarity implies that $|\mU|=9\Omega_0$, where $\Omega_0$ is the (universal) Phillips coefficient $\Omega_0$ discussed above. On the other hand, the weak similarity solution implies that $|\mU_w|= C_0\omega^{-x-\frac{1}{2}}$, where $C_0$ is a non-universal coefficient that depends on initial conditions. Since the exponent is positive ($\approx 0.417$), the condition $|\mU_w|\ll 1$ will be violated at some point in the UV, and the weak regime will no longer hold. This occurs near the transition point $\omega^{+} \simeq C_0^{\frac{1}{x+1/2}}$. Hence, the weak similarity only holds as a finite-size transient state in the IR, ultimately shifting to the Phillips similarity.

\begin{figure}
		\includegraphics[width=0.49\textwidth]{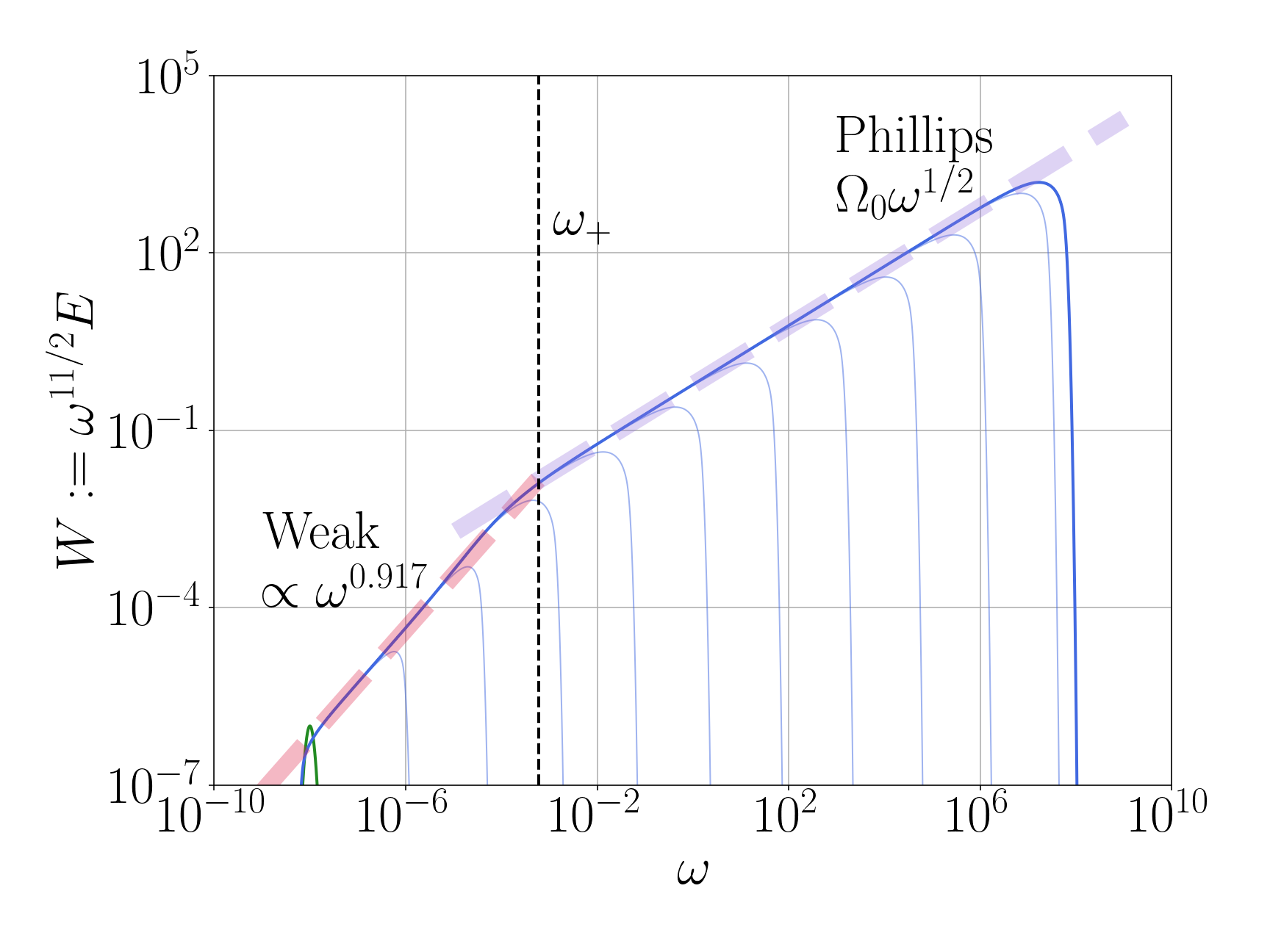}
		\includegraphics[width=0.49\textwidth]{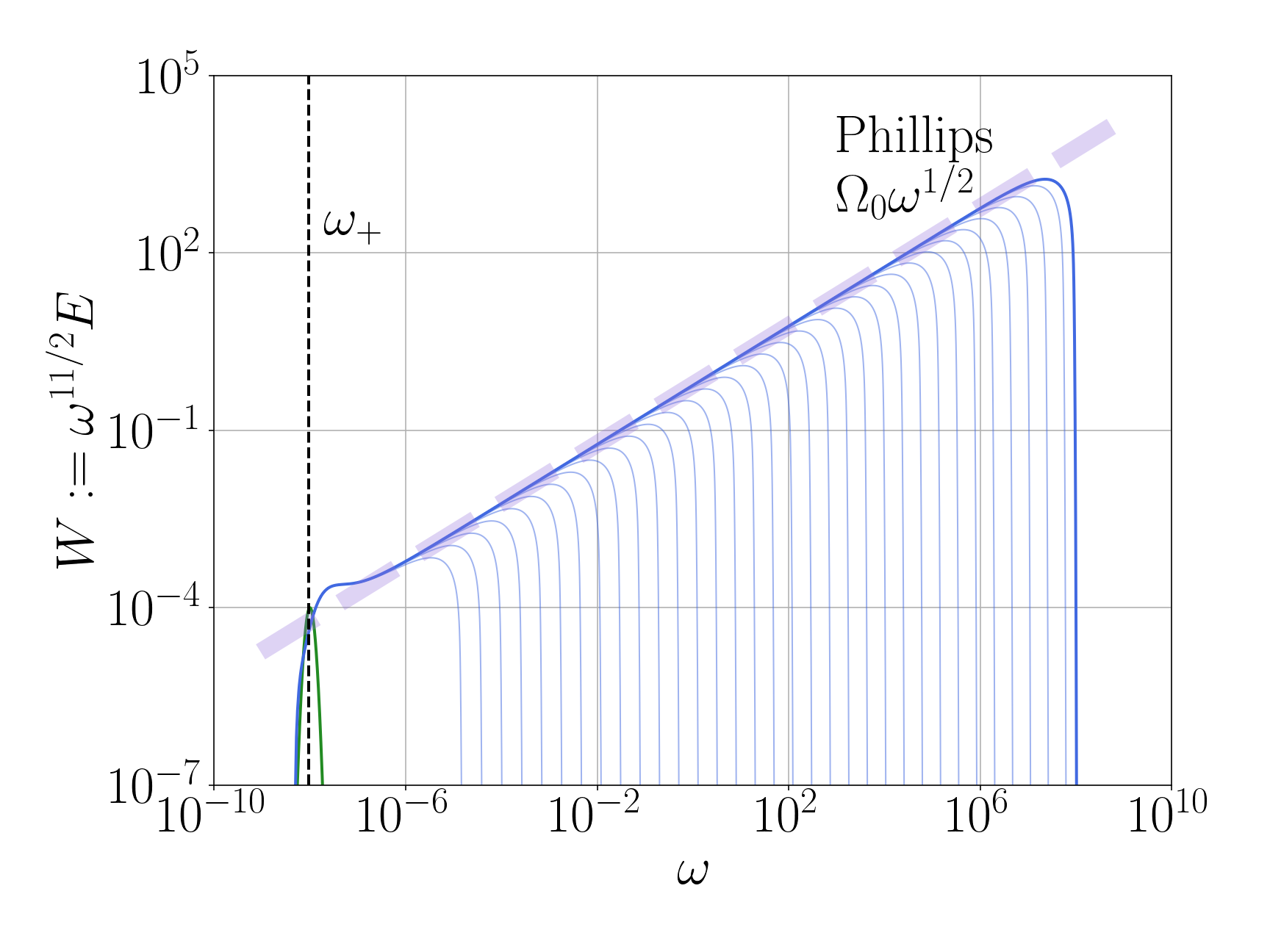}\\
		\includegraphics[width=0.49\textwidth]{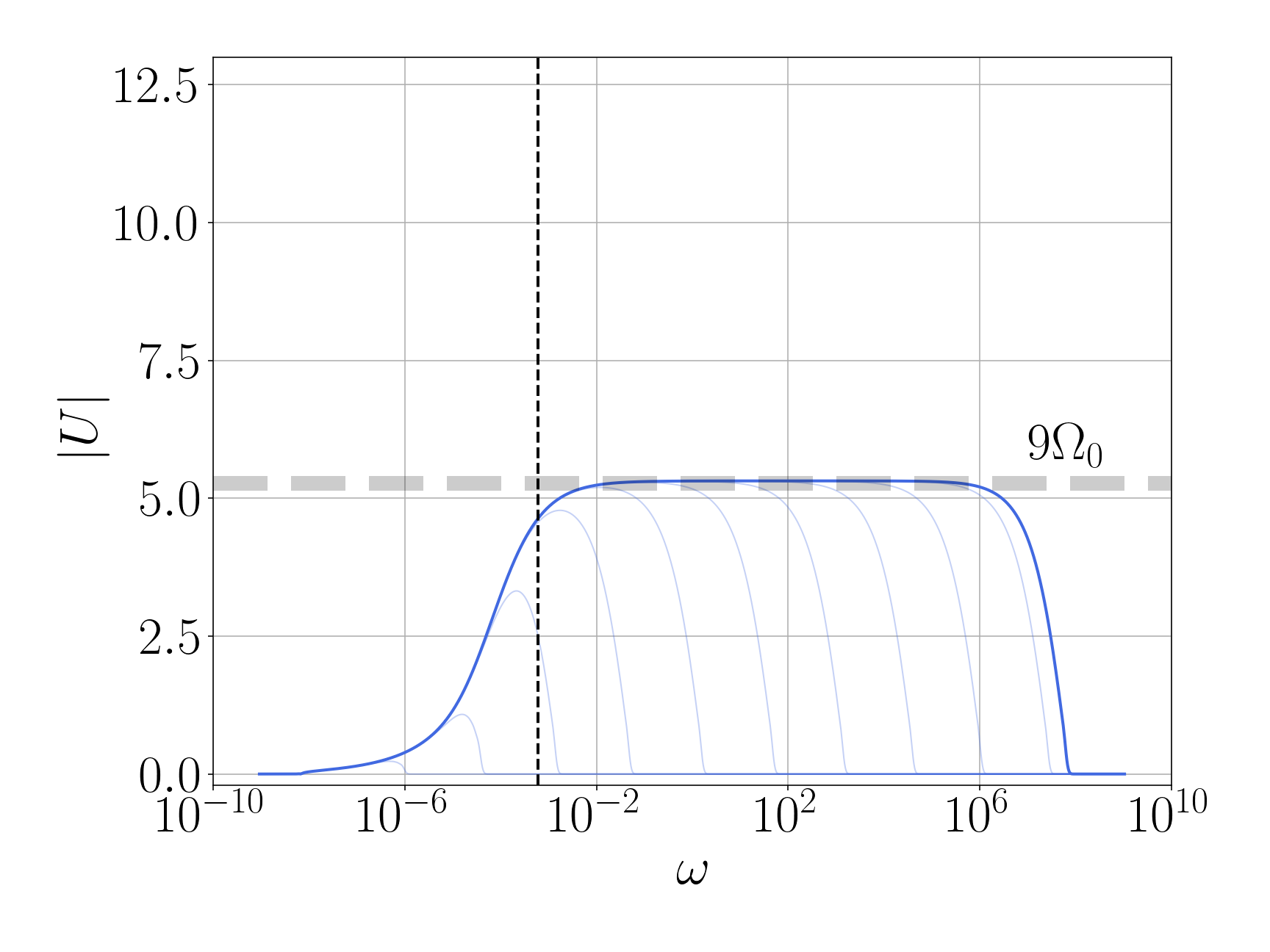}
		\includegraphics[width=0.49\textwidth]{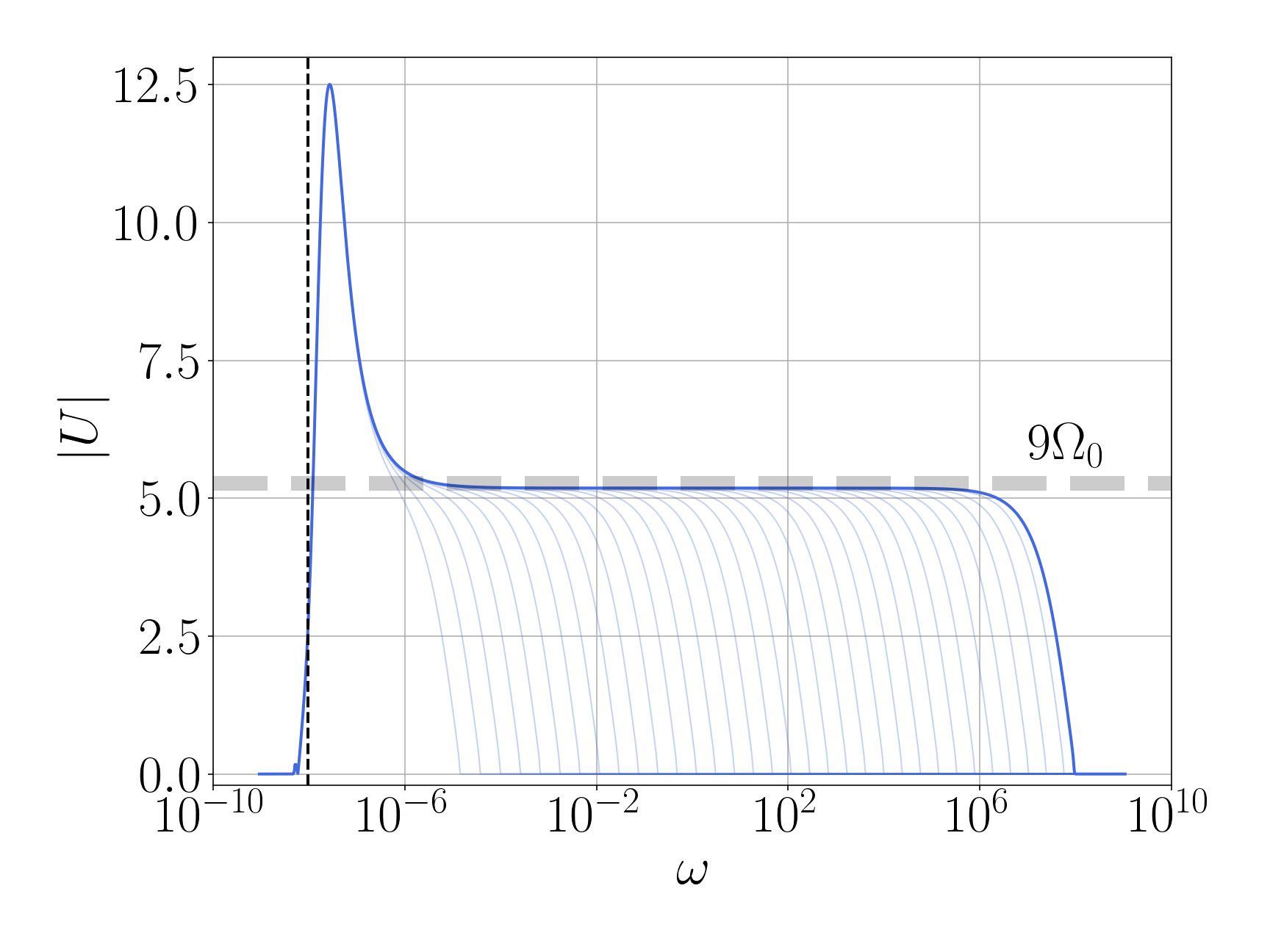}
		\caption{Convergence to Phillips in vorticity representation (top). The bottom panels show the corresponding factors $|U|$ determining the transitions. The initial profile is  centered at $\omega =10^{-8}$ and has  initial peak vorticity $ |W|_\infty \simeq 10^{-6} $ (left) and  $|W|_\infty\simeq 10^{-4}$ (right).}
	\label{fig:blowups}
	\end{figure}

The strong regime requires $|\mU| \gg 1$, but does not lead to any viable similarity solutions. 
A sharply peaked initial condition with $|\mU| \gg 1$ will therefore relax directly into the Phillips solution.  In other words,  the main consideration in defining the weak and strong regimes is the magnitude of the solution $W=\omega^{11/2}E$ with respect to the curve $\Omega_0 \omega^{1/2}$. This is  illustrated in the numerical simulations shown Fig.~\ref{fig:blowups}, which take $\phi=\pi/2$.

If the initial conditions lie well below this curve, the solution approximately obeys the weak \xdam\, with $K\approx K_W$. Over a range of frequencies and times it will approximately obey the weak self-similar ansatz \eqref{eq:ansatz} and evolve according to the anomalous scaling exponent $x=-0.917$. Upon reaching the  transition frequency $ \omega^{+}$ the solution becomes  of the same order as  $\Omega_0 \omega^{1/2}$ and the generic \xdam\, becomes relevant:  the solution then converges to the Phillips similarity. In the opposite limit in which the initial conditions are larger than $\Omega_0 \omega^{1/2}$, we may consider the strong regime in which $K\approx K_S$.  As discussed above,  this regime only supports diffusive solutions, and those relax towards the  Phillip similarity as $\omega$ increases towards the UV. 

\cb

\subsection{Bifurcation of blowups} 
Let us now discuss  the role of $\phi$.  As $\phi$ is varied, periodic oscillations appear around the Phillips solutions, which grow more pronounced as $\phi$ gets closer to $\pi$.
To characterize these oscillations, we appeal to the  the Dombre-Gilson viewpoint, which interprets  self-similar blowups as  solitonic waves propagating at constant speed in renormalized time $\tau = -\log(t_*-t)$ across the $\ln \omega$ range \cite{dombre1998intermittency,mailybaev2012computation,mailybaev2012renorm,campolina2025non}. 
To this end, instead of considering the self-similar ansatz  \eqref{eq:ansatz}, let us  first rescale the vorticity dynamics   using the  log-variables $\kappa,\tau$, 
\begin{equation}
	 \label{eq:DGrescaling}
	 \tilde W(\kappa,\tau) = (t_* - t)^{1/2} \omega^{11/2}E(\omega,t)   , \quad \text{with }\quad \kappa = \ln \omega,\quad \tau = -\log(t_* - t).
\end{equation}
Under the (DG) rescaling \eqref{eq:DGrescaling},  the rescaled vorticity evolves as 
\be
	\label{eq:DG}
	\partial_\tau \tilde W(\kappa,\tau)= \dfrac{-\tilde W}{2} +D_{-11/2}D_{-9/2}\left( \dfrac{\tilde W^4 D_{17/2}D_{19/2}\tilde W^{-1}}{\left|1-e^{j\phi}{e^{\tau/2-\kappa/2} D_{-19/2} \tilde W}\right|^2}\right),
\ee
with terms of  the homogenous  operators now reading $D_\alpha =  \frac{\partial}{\partial \kappa}  + \alpha \, \text{Id} $. The derivation of Eq.~\eqref{eq:DG} can be found in Appendix~\ref{ssec:weakODE}. From the DG dynamics 	\eqref{eq:DG}, one recovers the generic ODE given by Eq.~\eqref{eq:ODEPhil} upon identifying $\sigma = \kappa-\tau$ and solving for the solitonic profile  $\tilde W(\kappa,\tau) = \Omega(\sigma)$. In that sense, the similarity solutions  studied in \S \ref{sec:1} indeed map to solitonic waves in rescaled log-variables. 

\begin{figure}
\includegraphics[width=0.49\textwidth]{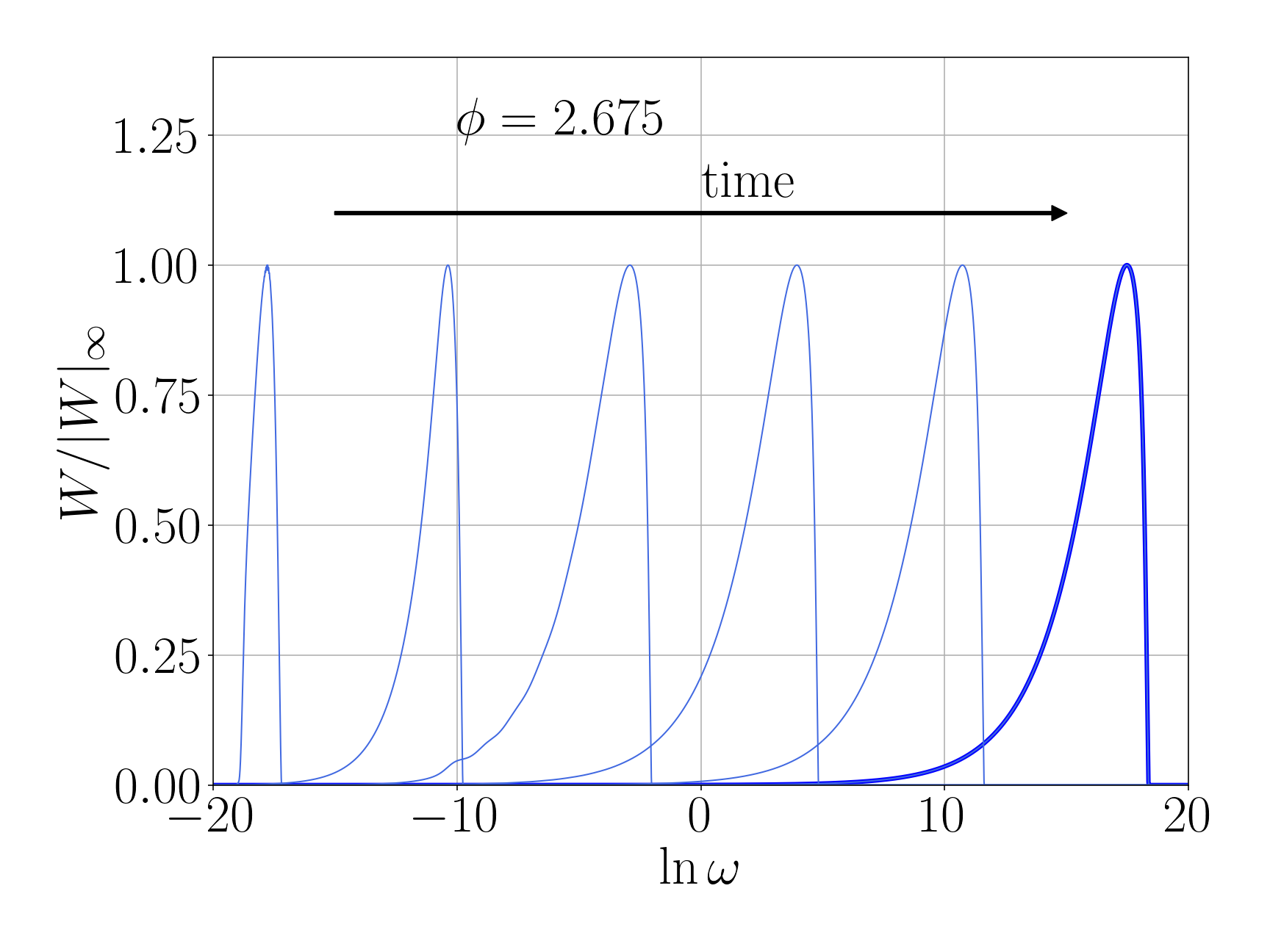}
\includegraphics[width=0.49\textwidth]{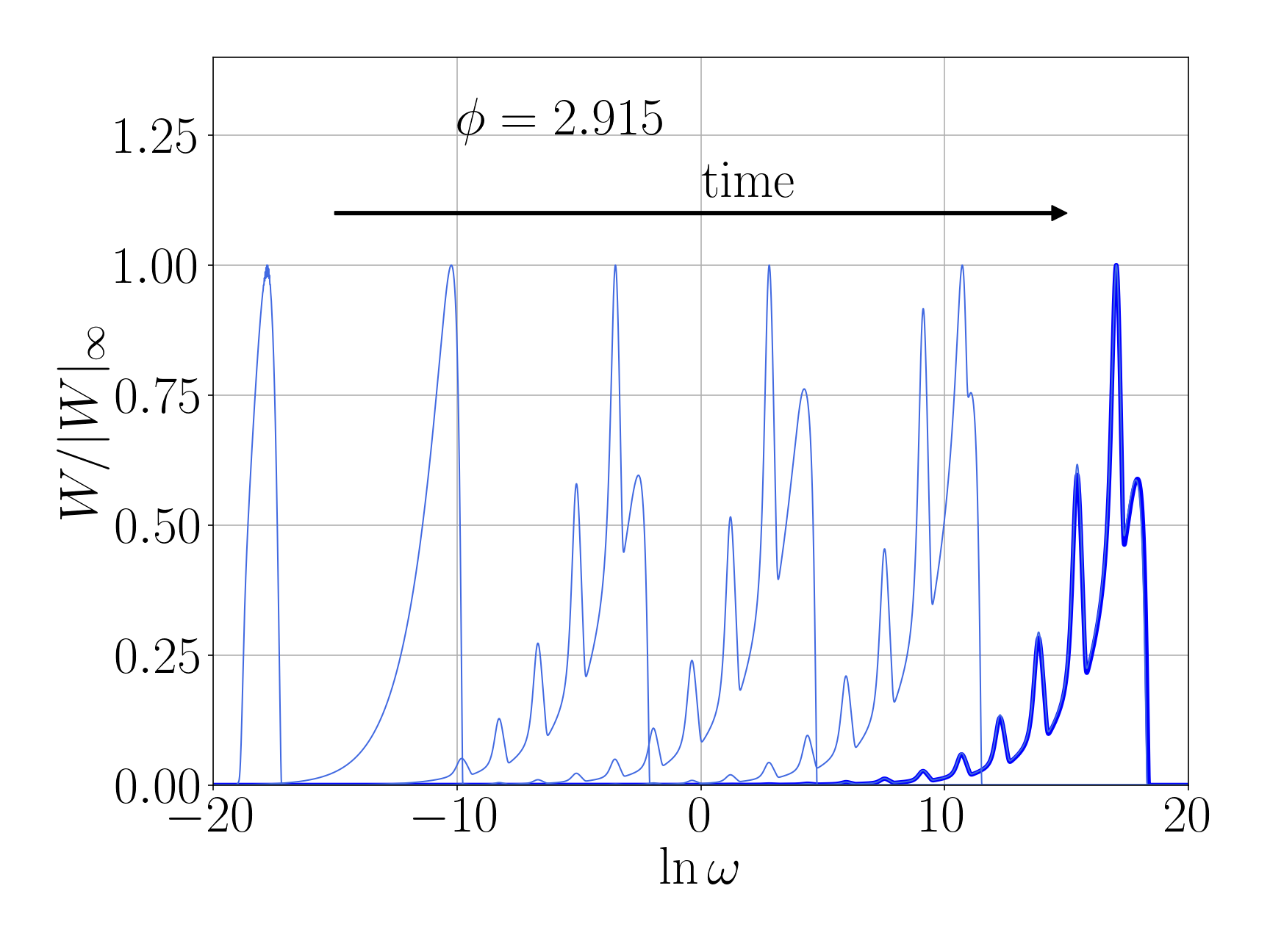}
\caption{Pure or periodic solitonic profiles observed for the DG rescaled vorticity $\tilde W(\kappa,\tau) = W/|W|_\infty$. Left: $\phi= 2.675$; Right: $\phi=2.915$}
\label{fig:5}
\end{figure}

Now, we may always rewrite $\tilde{W}$ as a function of $\sigma$, $\tilde{W}(\kappa,\tau)=\Omega(\sigma,\tau)$, and in general $\Omega$ will have additional $\tau$ dependence. A special class of so-called discrete self-similar solutions \cite{eggers2008role, mailybaev2013bifurcations} involve  periodic dependence in $\tau$ , $\Omega(\sigma, \tau+T)=\Omega(\sigma, \tau)$ for some period $T$. Our numerical simulations show that indeed these discrete self-similar solutions are the relevant solutions when $\phi >\phi_c\simeq 2.7$ ---See Fig. \ref{fig:5}. We also refer to this as the periodic blowup regime.

\begin{figure}
\includegraphics[width=0.49\textwidth]{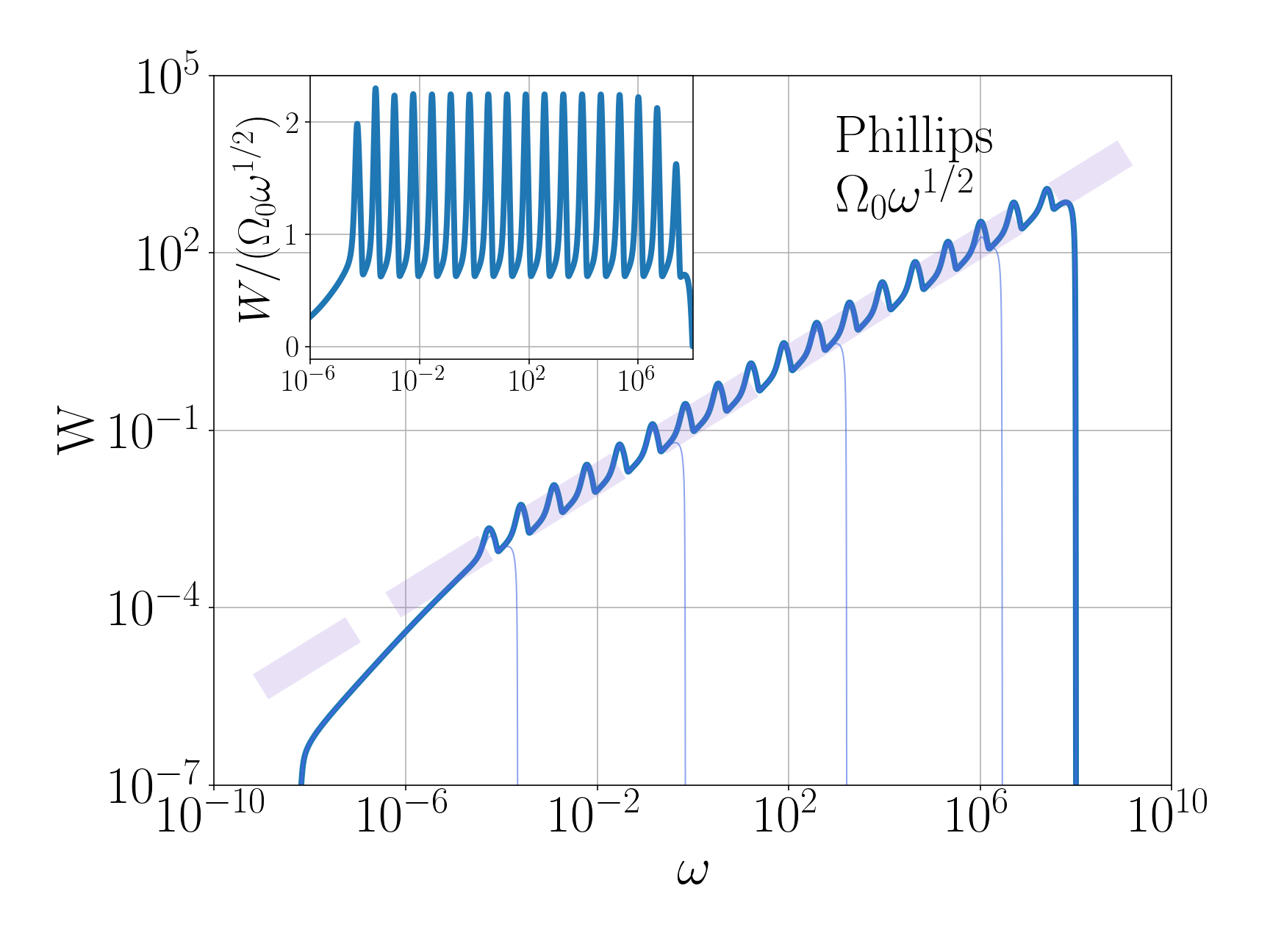}
\includegraphics[width=0.49\textwidth]{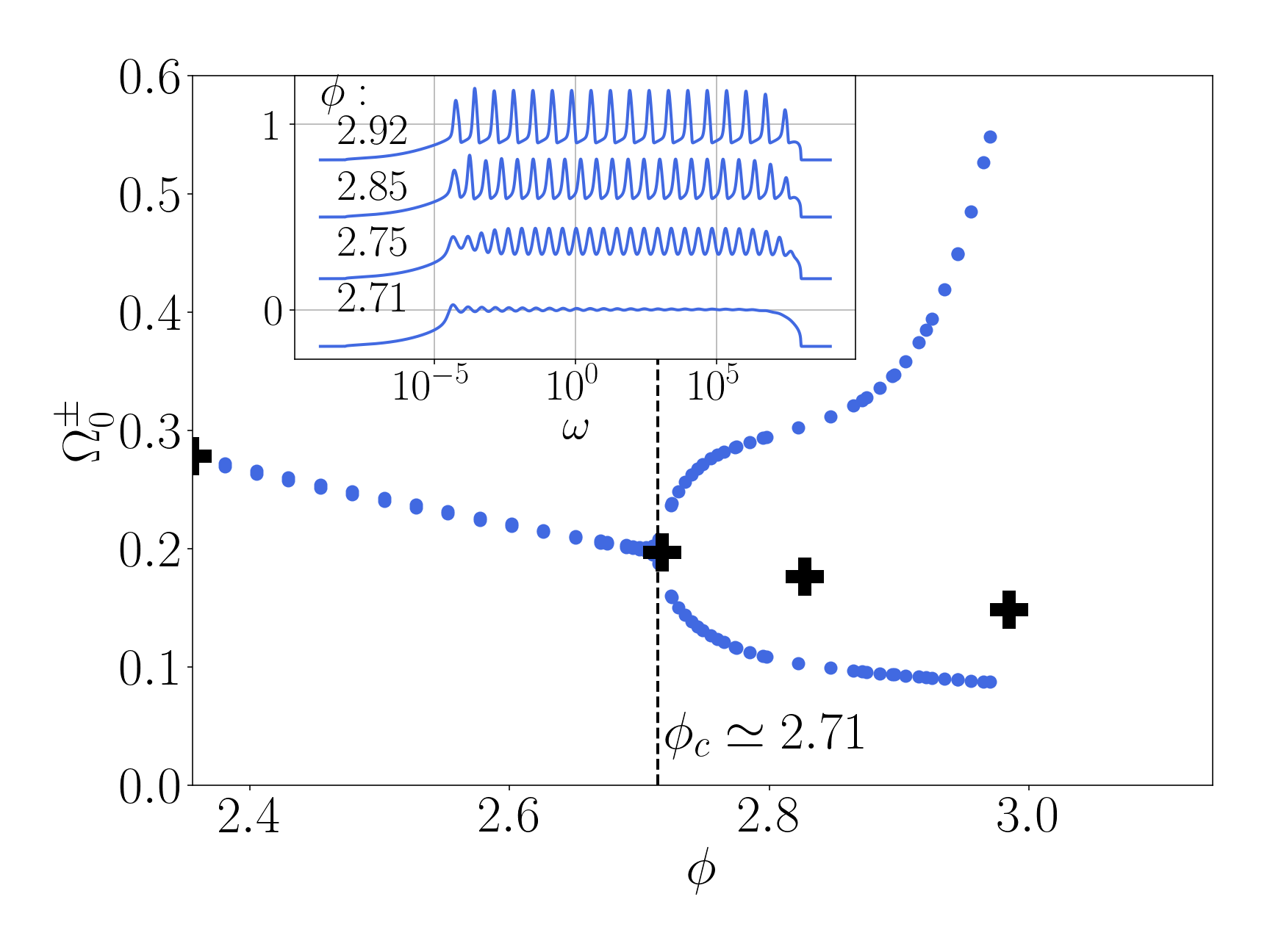}
\caption{Left: Vorticity spectrum for  $\phi=2.915$. The inset compensates the spectrum by the Phillips asymptotics to reveal the oscillatory part.  Right: Extrema $\Omega_0^{\pm}$ of the oscillations as a function of the phase-parameter $\phi$. The black crosses indicate the self-similar values found by the shooting method of \S~\ref{ssec:generic}. Inset shows the oscillatory patterns for various values of $\phi$.
}
\label{fig:6}
\end{figure}

We note that similar to the  calculation of Eq.~\eqref{eq:bkm}, the discrete self-similar solutions have their supremum diverging at the universal rate $|W|_\infty \propto (t_*-t)^{-1/2}$. This provides a rationale for the DG rescaling \eqref{eq:DGrescaling}, which can be interpreted as the dynamical rescaling of the vorticity by its (diverging) supremum in the rescaled time $\tau \propto 2 \ln |W|_\infty$. In the self-similar regime $\Omega(\sigma)$ was asymptotically $\sim \Omega_0e^{\sigma/2}$ for $\sigma\ll 0$. In the discrete (or periodic) self-similar regime, $\Omega_0$ is replaced by a periodic single variable function $\Omega_0(\sigma-\tau)$, leading to the asymptotic behavior
\begin{align}
	\Omega(\sigma, \tau)\sim \Omega_0(\sigma-\tau) e^{\sigma/2}, ~ \ \ \ \ \ \sigma\ll 0~.
\end{align}
Such periodic behavior is illustrated in  the left panel of Fig.~\ref{fig:6}.

The right panel of  Fig.~\ref{fig:6} illustrates the transition from self-similar to periodic blowup through a bifurcation diagram plotting the extrema of $U/9$ as a function of $\phi$. In the continuous similarity regime, the scaling properties far from the UV front impose
$U/9 =\Omega_0$,  recovering the universal  constant $\Omega_0$ computed by the shooting method described in  \S~\ref{ssec:generic}.
The minimum and maximum $\Omega_0^{\pm}$  of $U/9$  then coincide.

 In the periodic blowup regime an oscillatory pattern emerges, tracked by the splitting of $\Omega_0^\pm$ as $\phi>\phi_c$. Both the maximum and minimum values of the asymptotic function $\Omega_0(\sigma-\tau)$ are plotted. The eigenvalue problem  associated to the ODE \eqref{eq:ODEPhil} still admits a non-trivial solution---the results for $\Omega_0$ are the plotted black crosses. However, it is not realized for  $\phi>\phi_c$ in our numerics, suggesting that the continuous similarity solution has a vanishingly small basin of attraction.


\section{Concluding remarks}
\label{sec:4}
To summarize, the \xdam\, provides an example  of  kinetic wave dynamics in which the blowup scenario is determined by critical balance. In a sense, the emergence of a critical balance scaling is natural: In order to have a self-similar solution the two terms in the denominator (the $1$ and the $U$) must have the same scaling. As $U$ is the strength of the nonlinearity, this sets a critical balance condition.
However, this feature is not trivial, as the UV emergence of the Phillips spectrum in the \xdam\, is fragile in the steady state, being   restricted to exactly $\phi=\pi$. In the time-dependent setting, critical balance sets the UV behavior for any value  of the phase parameter $\phi \in [0,\pi)$.
We also emphasize that the critical balance at play in the \xdam\, does not require any ad-hoc linear damping mechanism--- 
at variance  for example with the  models studied in \cite{zakharov2012generalized,badulin2020phillips}.

Besides, the specific scenario for blowup depends  on the phase parameter. We have identified a critical phase  $\phi_c \simeq 2.7$, above which  periodic modulations emerge under  the Phillips envelope. 
As $\phi \to \pi$, the smooth periodic pattern transforms into a chain of pulses, and the limit $\phi=\pi$ seems ill-defined, giving rise to increasingly large fluctuations of $U$.  Note that such fluctuations were not encountered leading up to the steady-state Phillips behavior depicted in Fig.~\ref{fig:1}.
Due to the combination of  small-amplitude initial conditions and large viscous damping,
only the weak similarity regime appeared in the early-time regime of our numerics.
 Had the onset of dissipation been shifted deeper into the UV, or had the initial conditions been greater, these fluctuations would have been encountered;  this is another difficulty for realizing the steady-state Phillips behavior.
To our knowledge this is the first example of a non self-similar blowup in the context of the kinetic theory of waves. It is natural to wonder whether such a feature appears in  the more general kinetic equations, ones that do not involve the differential approximation.
 
\cm
This perspective is not restricted to the kinetics of gravity waves. While  this paper  focused on dispersion relation and interaction scaling appropriate for gravity waves, \xdam, as presented in \cite{rosenhaus2024strong}, is more general. \cb As detailed in Appendix \ref{Appendix A}, the general \xdam\, may have arbitrary values for the spatial dimension $d$ and the scaling exponents $\alpha$ and $\beta$ associated with the dispersion relation and interaction, respectively.
As long as $\beta/\alpha>3$, the stationary states are qualitatively similar to the gravity wave case, and similarly involve a transition from a weak direct cascade in the IR to a strong or Phillips direct cascade in the UV. However, the qualitative description of the blowup behavior involves a stronger condition on $\beta/\alpha$. The scenario involving a transition from weak to Phillips scaling depended on the anomalous exponent $-x\approx 0.917$ being greater than the corresponding Phillips exponent $1/2$ (see Fig.~\ref{fig:blowups}). For instance at $d/\alpha=4$, this requires the stronger condition $\beta/\alpha\gtrsim 3.67$. Below this range, the system may have weak anomalous scaling arbitrarily far in the UV and need not transition to Phillips.
For much smaller parameter values, the qualitative situation is inverted. 
\cm
In this case, relevant for the  nonlinear Schr\"odinger equation (Gross-Pitaevskii model)  \cite{Nowak:2011sk,B15, Walz:2017ffj, Gaz19, FR2, RosenhausSchubring2025}, there is weak anomalous scaling for the inverse cascade \cite{thalabard2021inverse}, and the \xdam\, has strong and Phillips stationary states in the IR rather than the UV.  
\cb
It would be natural to hypothesize that likewise there is a transition to Phillips blowup behavior as the solution propagates into the IR, but our preliminary numerical results suggest that this may not be the case. We leave this to future work.

\noindent \textit{Acknowledgments:} 
VR and DS are supported by NSF grant 2209116.

\bibliographystyle{siam}
\bibliography{biblio}

\appendix
\section{From kinetic equation to \xdam}\label{Appendix A}
 
 We here introduce the differential model of \cite{rosenhaus2024strong} in the context of gravity waves. Following closely the notation of \cite{badulin2005self}, the Zakharov equation \cite{zakharov1968stability} for gravity waves at the interface of a fluid is
\begin{align}
i\partial_t b_k = \omega_k b_k + \frac{1}{2}\int \tilde{T}_{k123}\,b^*_1 b_2 b_3 \delta_{k+1-2-3}dk_{123}~.\label{eq:zakharov}
\end{align}
Here $\omega_k=\sqrt{g|k|}$ is the dispersion relation   and $\tilde{T}$ is a complicated function of the wave numbers \cite{Krasitskii_1994, Onorato,korotkevich2024non}. Both $\omega$ and $\tilde{T}$ scale homogeneously
\begin{align}
	\omega_{\lambda k}=\lambda^\alpha \omega_k,\quad \tilde{T}_{\lambda k\,\lambda 1\,\lambda 2\,\lambda 3}=\lambda^\beta \tilde{T}_{k123},\label{def:alphaBeta}
\end{align}
with $\alpha=1/2$, $\beta=3$, and spatial dimension $d=2$ in the gravity wave case.

$b$ is a complex field that is constructed from the height above the wave surface $\eta$ and the velocity potential at the surface $\psi$. This is related to observable quantities through the wave action density $n_k\equiv \langle  b^* _k b_k \rangle$. Up to small corrections at higher order in the wave steepness, the wave action can be related to the average of $\eta^2$ at an arbitrary point in space,
\begin{align}
	\int d^2k\, \omega_k n_k = \int d\omega E(\omega)=\langle\eta^2 \rangle.
\end{align}
Explicitly for gravity waves, the energy density $E(\omega)$ defined in the second line is
\begin{align}
	E=	2\pi\omega\frac{d|k|}{d\omega}|k|n_k=4\pi g^{-2}\omega^4n_k
	.
\end{align}

The usual wave kinetic equation may be written in terms of $n_k$,
\begin{align}
	\partial_t n_k \propto \int dk_{123}\,\delta\left(\omega_{k1;23}\right)\left|\tilde{T}_{k123}\right|^2 n_k n_1 n_2 n_3 \left(\frac{1}{n_k}+\frac{1}{n_1}-\frac{1}{n_2}-\frac{1}{n_3}\right).\label{eq:waveKineticWeak}
\end{align} 
This may be reduced to a differential approximation model \cite{Hasselmann1985}, for details see \cite{DyachenkoNewellPushkarevZakharov1992},
\begin{align}
	\omega^{\frac{d}{\alpha}-1}\partial_t n_k \propto \partial_\omega^2\left(\omega^{\frac{3d+2\beta}{\alpha}} n^4 \omega^2\partial_\omega^2 n^{-1}\right).\label{eq:damWeak n}
\end{align} 
This may be rewritten in terms of $E$ and an explicit dimensionful coefficient $C_1$, which is proportional to $g^{-4}$ in the gravity wave case,
\begin{align}
	\partial_t E = \omega\partial_\omega^2\left(C_1\omega^{\frac{2\beta-d}{\alpha}} E^4 \omega^2\partial_\omega^2 \left(\omega^{\frac{d}{\alpha}}E^{-1}\right)\right).\label{eq:damWeak E}
\end{align} 

The wave kinetic equation \eqref{eq:waveKineticWeak} and the corresponding differential approximation model \eqref{eq:damWeak E} only involve interactions at lowest non-trivial order in perturbation theory, and thus are only valid in the weakly interacting regime. The wave kinetic equation may be extended by taking into account certain \emph{large N} corrections which are summed to all orders in perturbation theory \cite{Berges2002,ScheppachBergesGasenzer2010,rosenhaus2024strong}. 
The corrections included are the most significant ones in an $N$-component extension of the Zakharov equation \eqref{eq:zakharov} in the limit as $N$ becomes large, and it is hoped that even in the single component case the inclusion of these corrections may capture qualitative aspects of the strongly interacting regime.

The net effect of including the large $N$ corrections renormalizes the interaction coefficient in the kinetic equation. Schematically,~\footnote{Strictly speaking this expression only holds for $\tilde{T}$ with factorizable momentum dependence, but this is taken as an assumption in the reduction to the differential approximation model.} the result takes the form
\begin{align}
\label{eq:A8}
\left|\tilde{T}_{k123}\right|^2\rightarrow \frac{\left|\tilde{T}_{k123}\right|^2}{\left|1-2\int dk_{45} \tilde{T}_{k425}\frac{n_4-n_5}{\omega_{k4;25}+i\epsilon}\right|^2}.
\end{align}

Reducing to a DAM in the manner of \cite{DyachenkoNewellPushkarevZakharov1992}, the difference $n_4-n_5$ becomes a derivative $\omega\partial_\omega n$, the coefficient $\tilde{T}$ introduces a factor of $\omega^{\beta/\alpha}$, and the integration over a single independent momentum introduces a factor $\omega^{d/\alpha}$. The remaining integration introduces a second constant $C_2$, which is proportional to $g^{-2}$ in the gravity wave case.\footnote{Note that $C_1$ has absorbed a factor of the small parameter $1/N$ in the large $N$ approach, but $C_2$ is of order $1$.} In total the DAM in terms of $E$ is modified to
\begin{align}
	\partial_t E = \omega\partial_\omega^2\left(\frac{C_1\omega^{\frac{2\beta-d}{\alpha}} E^4 \omega^2\partial_\omega^2 \left(\omega^{\frac{d}{\alpha}}E^{-1}\right)}{\left|1-C_2\omega^{\frac{d+\beta-\alpha}{\alpha}}\omega\partial_\omega\left(\omega^{-\frac{d}{\alpha}}E\right)\right|^2}\right).\label{eq:dam}
\end{align} 
This is the $N$-DAM investigated in the main body, except from the presence of the additional parameters $C_1$ and $C_2$. Here $C_1$ is real and positive, but $C_2$ is complex in general, and may be written as $C_2=|C_2|e^{i\phi}$. We may eliminate the explicit appearance of $C_1, |C_2|$ by working with rescaled quantities
$$\tilde{N}=|C_2|N,\qquad \tilde{t}=\frac{C_1}{|C_2|^2}t,\qquad \tilde{P}=\frac{|C_2|^2}{C_1}P.$$
In the formulation of the $N$-DAM used throughout the text, we use these rescaled quantities exclusively, and omit the tildes.
\cm
This  derivation highlights that the phase  $\phi$ is not introduced \emph{ad hoc}, but instead emerges as the phase of a complex constant $C_2$ generated by the renormalized interaction coefficient in Eq.~\eqref{eq:A8}.
\cb

\section{Technical details}
\label{sec:technical}
\subsection{Derivation of  the weak ODE~\eqref{eq:ODEweak}}
\label{ssec:weakODE}
To derive the ODE \eqref{eq:ODEweak}, we use the vorticity defined in Eq.~\eqref{eq:bkm} as
\be 
	W(\omega,t) = \omega^{11/2} E(\omega,t),
\label{eq:W}
\ee
 whose dynamics
in the absence of forcing and dissipation is derived from the \xdam\, \eqref{eq:RSDAM} as
\be
	\label{eq:WP}
	\partial_t W (\omega,t) + \omega^{11/2}\partial_\omega P =  0
\ee
where $P$ is prescribed by Eq.~\eqref{eq:RSDAM} as
\be
	\label{eq:Kbis}
	 P =- \omega^2\partial_\omega\left(\omega^{-1} K \right),\quad
  	K(\omega,t)= \frac{\omega^{10} E^{4} \partial^2_{\omega \omega} (\omega^4 E^{-1})}{|1-e^{j\phi}U |^2},\quad U=\omega^{10}\partial_\omega \left(E\omega^{-4}\right).
\ee

Setting $U=0$ and using the expresion \eqref{eq:W} yields the weak version of the flux term as   
\be
	\begin{split}
	\omega^{11/2}\partial_\omega P & = -\omega^{11/2}\partial_\omega\left( \omega^{2} \partial_\omega \left(\omega^{-13}W^4 \partial_{\omega\omega} \left( \omega^{19/2} W^{-1}\right)\right)\right)\\
	& =  - \tilde D_{-9/2}  \tilde D_{-11/2} \left({W^4    \tilde D_{17/2}\tilde D_{19/2}W^{-1}}\right),
	\end{split}
	\label{eq:rhsP}
\ee
where the second line is obtained by successively substituting --- from right to left --- the derivatives by the operators $ D_\alpha = \omega\partial_\omega + \alpha \, Id$, observing in particular that
\be 	
	\partial_\omega \left(\omega^\alpha f\right) =  \omega^{\alpha-1}  D_\alpha f.
\ee
We now use the ansatz \eqref{eq:ansatz}, which in terms of $W$ simplifies to
\be
	W(\omega,t) = (t_*-t)^{bx} \Omega(\eta),\quad\eta = (t_*-t)^b\omega,\quad b(x)={-\dfrac{1}{2x}}.
	\label{eq:Wan}
\ee
Inserting 	the ansatz~\eqref{eq:W} into Eq.~\eqref{eq:WP} yields
for the first term
\be
	\partial_t W = 2x (t_*-t)^{-3/2}\left(x \Omega+  \eta \dfrac{\de}{\de\eta}\Omega \right) = 2x (t_*-t)^{-3/2}D_x \Omega,
	\label{eq:WP1}
\ee
where the $D_\alpha$ operator is now prescribed by Eq.~\eqref{eq:Dalpha}.
For the second term, we get
\be
	\omega^{11/2}\partial_\omega P = 
  -(t_*-t)^{-3/2} D_{-9/2}  D_{-11/2} \left(  \Omega^4   D_{17/2}D_{19/2}\Omega^{-1}\right).
	\label{eq:WP2}
\ee
As a result of 	Eq.\eqref{eq:WP}, Eq.~\eqref{eq:WP1} and \eqref{eq:WP2} sum to $0$.  Cancelling the time prefactor $(t_*-t)^{-3/2}$  then yields the weak ODE \eqref{eq:ODEweak}.

\subsection{Derivation of the DG dynamics \eqref{eq:DG}}
\label{ssec:weakODE}
In terms of the vorticity $W$ given by Eq.~\eqref{eq:W}, the term $U$ in Eq.~\eqref{eq:Kbis} reads 
\be
	U=\omega^{-1/2 } D_{-19/2}W.
\ee
Repeating the calculations of \S~\ref{ssec:weakODE}, the vorticity dynamics \eqref{eq:WP} recasts as 
\be
	\label{eq:DG-recall}
	\partial_t W(\omega,\tau)=F[W],\quad F:= D_{-11/2}D_{-9/2}\left( \dfrac{ W^4  D_{17/2} D_{19/2}W^{-1}}{\left|1-e^{j\phi}{\omega^{-1/2} D_{-19/2}W}\right|^2}\right),
\ee
Now using Eq.~\eqref{eq:DGrescaling}, we   introduce the new variables $\kappa=\ln \omega$, and $\tau=-\ln(t_*-t)$. The dynamics of   the rescaled profile $\tilde W(\kappa,\tau) =e^{\tau/2} W$ is then  obtained as 
\be
	\partial_t \tilde W = -\dfrac{\tilde W}{2} +e^{-3\tau/2} F[W].
\ee
With the differential operator becoming $D_\alpha \to  \partial_\kappa +\alpha \text{Id}$,
observing that 
\be
	F[W] = F[e^{\tau/2}\tilde W] = e^{3\tau/2}   D_{-11/2}  D_{-9/2}\left( \dfrac{ \tilde W^4  D_{17/2}  D_{19/2}\tilde W^{-1}}{\left|1-e^{j\phi}{e^{\tau/2-\kappa/2}  D_{-19/2}\tilde W}\right|^2}\right)
\ee
 yields Eq.~\eqref{eq:ODEweak}.

\end{document}